\documentclass[11pt,twoside]{article}

\usepackage{asp2006}
\usepackage{graphicx}

\begin{document}
\title{Variable X-ray Absorption in mini-BAL QSOs}   

\author{Margherita Giustini,\altaffilmark{1,2,3} Massimo Cappi,\altaffilmark{3} George Chartas,\altaffilmark{1} Michael Eracleous,\altaffilmark{1} Giorgio G.C. Palumbo,\altaffilmark{2} and Cristian Vignali,\altaffilmark{2}} 

\altaffiltext{1}{Department of Astronomy \& Astrophysics, Penn State University, University Park, PA 16802} 
\altaffiltext{2}{Dipartimento di Astronomia, Universit\`a di Bologna, via Ranzani 1, 40127 Bologna, Italy}
\altaffiltext{3}{INAF/Istituto di Astrofisica Spaziale e Fisica cosmica, via Gobetti 101, 40138 Bologna, Italy}

\begin{abstract} 
We present the results of X-ray spectral analysis of two mini-BAL QSOs, PG~1126-041 and PG~1351+640, aimed at getting insights into the physics of quasar outflows. 
We find strong X-ray spectral variability on timescales of years. These variations can be well reproduced by variations of physical properties as the covering factor and column density of ionized gas along the line of sight, compatible with radiatively-driven accretion disk wind models. 
\end{abstract}

\section{Introduction}  
X-ray observations are crucial in getting insights into the physical mechanism responsible for launching and accelerating quasar outflows. 

For example, radiatively-driven accretion disk wind models predict the presence of high column densities of X-ray absorbing gas which prevents the overionization of the UV-driven wind, allowing for its effective acceleration \citep[see e.g.][]{PR04}. Furthermore, the physical properties of this X-ray absorbing gas are predicted to be variable with time \citep[see e.g.][]{SC09}.
However, high signal-to-noise ratio X-ray observations of BAL QSOs are rare, because of their known ``X-ray weakness'' -- e.g. \citeauthor{GAL06}~\citeyear{GAL06}; but see \citeauthor{GIU08}~\citeyear{GIU08}. 

We present the results of multiple XMM-\textit{Newton} observations of two low redshift mini-BAL QSOs, PG~1126-041 (z=0.06) and PG~1351+640 (z=0.09).

\begin{figure}
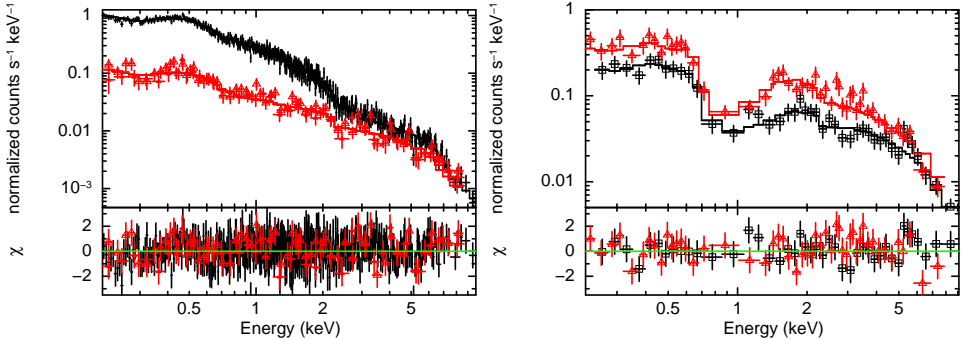

\includegraphics[scale=0.245, angle=-90]{giustinim_f1.ps}
\includegraphics[scale=0.245, angle=-90]{giustinim_f2.ps}
\caption{Left: EPIC-pn spectra of PG~1351+640 taken in December 2004 (black dots) and in June 2008 (red triangles). Right: EPIC-pn spectra of PG~1126-041 taken in June 2008 (black squares) and in December 2008 (red triangles) epochs. Bottom: $\chi^2$ residuals in units of $\sigma$.\label{FIG1}}
\end{figure}
\section{Data Analysis and Results}
We analyzed five EPIC datasets, three for PG~1351+640 (observed in 2004/12/31, 2008/06/06 and 2008/06/08) and two for PG~1126-041 (observed in 2008/06/15 and 2008/12/13).
We reduced all the datasets following standard XMM-\textit{Newton} analysis threads. \\
\textbf{PG~1351+640:}\  The source flux dropped by a factor of $\sim$10 in the soft ($E < 1$~keV) band, while remaining constant at $E > 7$~keV during 4 years. No changes are detected during the two days elapsed between the two observations taken in June 2008, so we analyzed them together. The best fit model ($\chi^2/$d.o.f.=875/877, Fig.\ref{FIG1}, left) is a powerlaw continuum with $\Gamma\sim 2.3$ absorbed by two layers of ionized gas; one is totally covering the continuum source, has a column density $N_H\sim 1.5 \times 10^{21}$~cm$^{-2}$ and an ionization parameter $\xi\sim 20$~erg~cm~s$^{-1}$ in both states. The other ionized absorber has an ionization parameter $\xi\sim 300$~erg~cm~s$^{-1}$ and is covering ~50\%(~90\%) of the source, with a column density $N_H\sim 9(1)\times 10^{23}$~cm$^{-2}$  in the 2004(2008) observation.\\
\textbf{PG~1126-041:}\ The 0.2-10 keV source flux increased by a factor of $\sim$60\% in the six months elapsed between the two observations presented here. The best fit model ($\chi^2/$d.o.f.=96/102, Fig.\ref{FIG1}, right) is a powerlaw continuum with $\Gamma\sim 2.2$ absorbed by a layer of ionized gas covering $\sim 95$\% of the source, with a column density $N_H\sim 8(5) \times 10^{22}$~cm$^{-2}$ and an ionization parameter $\xi\sim 30(40)$~erg~cm~s$^{-1}$ in the 2008 June(December) observation.

\section{Conclusions}
Our observations are consistent with radiatively-driven accretion disk wind models in that they find high column densities of X-ray absorbing gas along the line of sight, and strong variability in the physical properties of these ionized absorbers. Future, deeper X-ray observations of quasars which show powerful UV-driven winds are needed to confirm this scenario: for example, the adopted models predict the presence of strong iron absorption lines which cannot be detected with the present moderate signal-to-noise ratio spectra.

\acknowledgements 
We acknowledge financial support from NASA grant NNX08AB71G and ASI contract I/088/06/0.

\end{document}